\documentclass[%
 reprint,
 amsmath,amssymb,
 aps,
]{revtex4-1}

\usepackage{graphicx}
\usepackage{dcolumn}
\usepackage{bm}
\usepackage{lineno,hyperref}
\usepackage{xspace}
\modulolinenumbers[5]
\usepackage{bm}
\usepackage{amssymb}
\usepackage{hyperref}
\usepackage{amsmath}

\newcommand{\snnt}[1]{\ensuremath{\sqrt{s_{\rm NN}} = #1 \text{\,TeV}}\xspace}
\newcommand{\pt}{\ensuremath{p_{\rm{T}}}\xspace}

\begin{document}

\preprint{APS/123-QED}

\title{Energy density and path-length dependence of the fractional momentum loss in heavy-ion collisions at $\sqrt{s_{\rm NN}}$ from 62.4 to 5020\,GeV}

\author{Antonio Ortiz}
 \email{antonio.ortiz@nucleares.unam.mx}
\author{Omar V\'azquez}%
\affiliation{%
 Instituto de Ciencias Nucleares, Universidad Nacional Aut\'onoma de M\'exico, \\ Apartado Postal 70-543,
M\'exico Distrito Federal 04510, M\'exico
}%

\date{\today}

\begin{abstract}
In this work a study of the fractional momentum loss ($S_{\rm loss}$) as a function of the characteristic path-length ($L$) and the Bjorken energy density times the equilibration time ($\epsilon_{\rm Bj}\tau_{0}$) for heavy-ion collisions at different $\sqrt{s_{\rm NN}}$ is presented. The study has been conducted using inclusive charged particles from intermediate to large transverse momentum ($5<p_{\rm T}<20$\,GeV/$c$). Within uncertainties and for all the transverse momentum values which were explored, the fractional momentum loss is found to increase linearly with $({\epsilon_{\rm Bj}\tau_{0}})^{3/8}$$L$. For identified hadrons, albeit a smaller slope of $S_{\rm loss}$ vs. $({\epsilon_{\rm Bj}\tau_{0}})^{3/8}$$L$ is observed for (anti)protons at $\pt^{\rm}=5$\,GeV/$c$, $S_{\rm loss}$ is also found to grow linearly with $L$. The behaviour of data could provide important information aimed to understand the parton energy loss mechanism in heavy-ion collisions and some insight into the expected effect for small systems.
\begin{description}
\item[PACS numbers]
25.75.-q, 12.38.Mh
\end{description}
\end{abstract}

\pacs{Valid PACS appear here}
\maketitle


\section{\label{sec:level1}Introduction}

Ultrarelativistic heavy-ion collisions allow the study of a new form of matter featured by deconfinement. In $\sqrt{s_{\rm NN}}=200$\,GeV Au-Au collisions, experiments at the RHIC claimed the discovery of a QGP which behaved like a perfect fluid, and not as the expected gas~\cite{Adams:2005dq,Back:2004je,Arsene:2004fa,Adcox:2004mh}.  This strongly interacting Quark-Gluon Plasma was characterized by a strong collective flow  and jet quenching~\cite{Gyulassy:2004zy,Dainese:2004te}. These results were later confirmed and further extended in \snnt{2.76} Pb-Pb collisions  at the LHC~\cite{Bala:2016hlf}.

The study of the propagation of a hard probe through the medium offers the possibility to determine the properties of the QGP. Experimentally, the medium effects are extracted by comparing measurements on pp and A-A collisions, for example, particle production at large transverse momentum (\pt) in A-A (${\rm d}^2N_{\rm AA}/{\rm d}y{\rm d}\pt$) to that in pp (${\rm d}^2N_{\rm pp}/{\rm d}y{\rm d}\pt$). Commonly, the nuclear modification factor is used to quantify the changes:
\begin{equation}
R_{\rm AA}=\frac{{\rm d}^2N_{\rm AA}/{\rm d}y{\rm d}\pt}{\langle N_{\rm coll} \rangle{\rm d}^2N_{\rm pp}/{\rm d}y{\rm d}\pt}
\end{equation} 
where $\langle N_{\rm coll} \rangle$, usually obtained using Glauber simulations~\cite{Loizides:2014vua}, is the average number of binary collisions occurring within the same A-A interaction. Clearly, in the absence of medium effects, i.e. superposition of nucleon-nucleon collisions, $R_{\rm AA}$ would be one. Several measurements of $R_{\rm AA}$ for different $\sqrt{s_{\rm NN}}$~\cite{Adler:2003au,Adams:2003kv,Aamodt:2010jd,CMS:2012aa,Aad:2015wga} support the formation of a dense partonic medium in heavy nuclei collisions where hard partons lose energy via a combination of both elastic and inelastic collisions with the constituents of QGP~\cite{Qin:2015srf}. However, in this work we used the inclusive charged particle suppression data to get an alternative estimate of the jet-quenching effects: the fractional momentum loss proposed by the PHENIX Collaboration~\cite{Adare:2015cua}. This because there are some effects which can affect our interpretation of energy loss from $R_{\rm AA}$ measurements. For instance, while energy loss increases with increasing $\sqrt{s_{\rm NN}}$ which would tend to decrease $R_{\rm AA}$, the pp production cross section of high $p_{\rm T}$ particle goes like: 
\begin{equation}
\frac{{\rm d}^{2}\sigma_{\rm pp}(p_{\rm T})}{{\rm d}y{\rm d}p_{\rm T}} \propto \frac{1}{{p}_{\rm T}^{n}}
\end{equation}
Therefore, a countervailing effect on $R_{\rm AA}$ is expected since the power $n$ decreases with increasing $\sqrt{s_{\rm NN}}$.

\begin{figure*}[t!]
\begin{center}
\includegraphics[keepaspectratio, width=2.0\columnwidth]{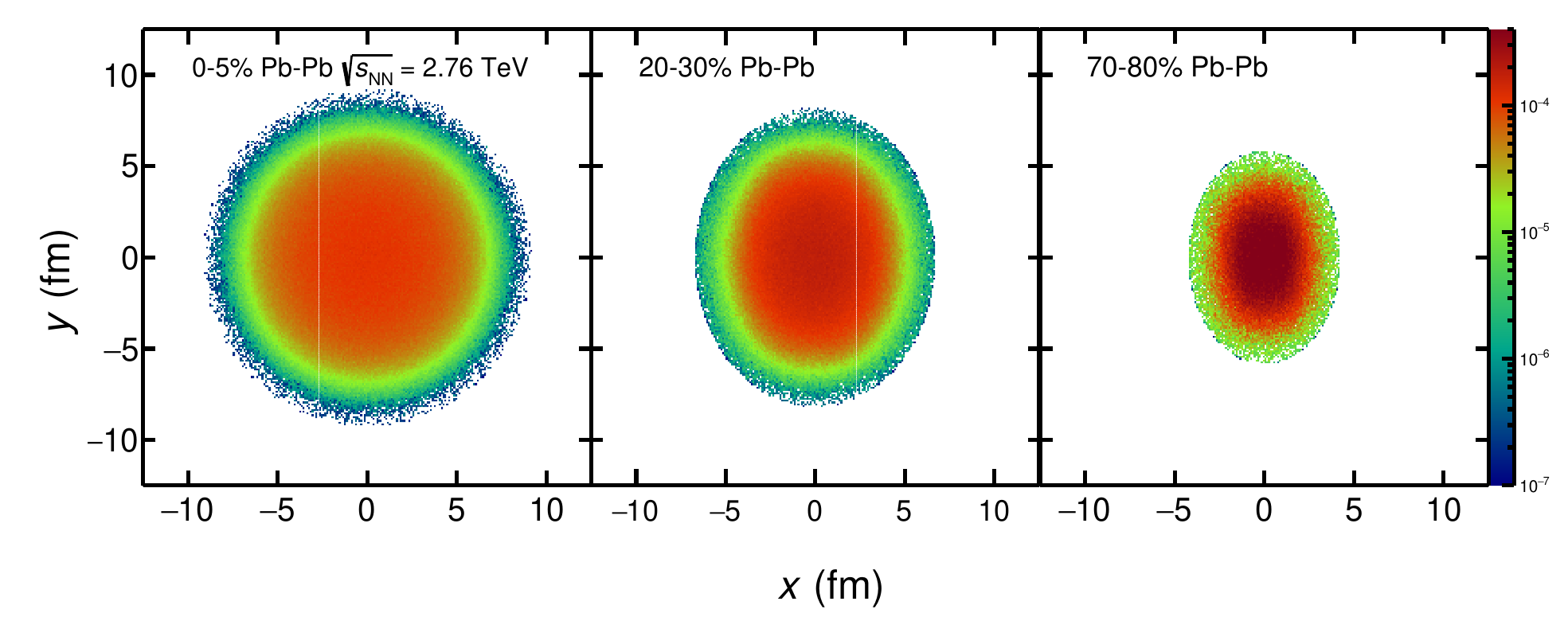}
\caption{\label{fig:1} (Color online) Number of participants distributions obtained from Glauber simulations for Pb-Pb collisions at $\sqrt{s_{\rm NN}}=2.76$\,TeV. Results for 0-5\% (left), 20-30\% (middle) and 70-80\% (right) are displayed.} 
\end{center}
\end{figure*}

As discussed in~\cite{Baier:2001yt}, at large transverse momenta, yields are mainly suppressed by means of medium induced gluon radiation accompanying multiple scattering. To model energy loss effects, the authors proposed to convolute the vacuum (pp) production cross section of the particle with energy $p_{\rm T} + \varepsilon$ with the distribution $D(\varepsilon)$ that describes specifically the additional energy loss $\varepsilon$ due to medium induced gluon radiation in the final state. Thus, the minimum bias (centrality integrated) heavy-ion production cross section reads:
\begin{equation}
\frac{{\rm d}^{2}\sigma_{\rm AA}(p_{\rm T})}{{\rm d}y{\rm d}p_{\rm T}} \propto \int_{0}^{\infty} {\rm d}\varepsilon D(\varepsilon) \frac{{\rm d}^{2}\sigma_{\rm pp}(p_{\rm T}+\varepsilon)}{{\rm d}y{\rm d}p_{\rm T}} 
\end{equation}
Where $\varepsilon$ is characterized by the scale $\omega_{\rm c}=\hat{q}L^{2}/2$ being $\hat{q}$ the transport coefficient which controls the medium dependence of the energy loss and $L$ the medium length. The quenching effect can be modelled by the substitution:
\begin{equation}
\frac{{\rm d}^{2}\sigma_{\rm AA}(p_{\rm T})}{{\rm d}y{\rm d}p_{\rm T}} =  \frac{{\rm d}^{2}\sigma_{\rm pp}(p_{\rm T}+\delta_{p_{\rm T}})}{{\rm d}y{\rm d}p_{\rm T}} 
\end{equation} 
Taking into account the interplay between the energy loss and the pp cross section fall-off, the $p_{\rm T}$ dependent expression for the shift goes like:
\begin{equation}
\delta_{p_{\rm T}} \approx (p_{\rm T}\omega_{\rm c})^{1/2} 
\end{equation} 
Considering $\omega_{\rm c}=\hat{q}L^{2}/2$ and that the ideal estimate from pQCD calculations yields to $\hat{q}\propto\epsilon^{3/4}$~\cite{Baier:2002tc}, being $\epsilon$ the energy density of the system. One would expect:
\begin{equation}
\delta_{p_{\rm T}} \approx p_{\rm T}^{1/2} \epsilon^{3/8}L
\end{equation} 
Clearly, $\delta_{p_{\rm T}}$ does not equal the mean medium induced energy loss, $\Delta E \propto L^{2}$. It has been shown that $\delta_{p_{\rm T}}$ can be related with the fractional momentum loss~\cite{Tywoniuk:2017jbz} and that a linear relation between fractional momentum loss and $\epsilon^{3/8}L$ is required in order to simultaneously describe the azimuthal anisotropies and $R_{\rm AA}$ at high $p_{\rm T}$~\cite{Christiansen:2013hya}. Moreover, a recent work has also exploited these ideas in order to explain scaling properties of $R_{\rm AA}$~\cite{Arleo:2017ntr}.

Inspired by recent data-driven studies, where the parton energy loss has been separately studied as a function of the Bjorken energy density~\cite{Bjorken:1982qr} times the formation time ($\epsilon_{\rm Bj}\tau_{0}$)~\cite{Adare:2015cua} and a characteristic path-length~\cite{Christiansen:2013hya,Christiansen:2016uaq}, in the present work other possibilities are explored. Namely, based on the preceding discussion the fractional momentum loss is studied as a function of $\epsilon_{\rm Bj}\tau_{0}$ and $L$. Where, for the estimation of the characteristic path length the different geometry for the trajectories have been taken into account. To this end, the ideas presented in~\cite{Dainese:2004te,Ayala:2009fe} were implemented. Namely, energy density distributions estimated with Glauber simulations~\cite{Loizides:2014vua} were considered as the distributions of the scattering centers. This allows to test the previously discussed energy loss model~\cite{Baier:2001yt} by means of the fractional momentum loss for several transverse momentum values and for the top energy reached at the LHC, \snnt{5.02}~\cite{Khachatryan:2016odn}. 

The paper is organized as follows, section 2 describes how the different quantities: path-length, Bjorken energy density and fractional momentum loss, were extracted from the data. The results and discussions are displayed in section 3 and the final remarks are presented in section 4.

\section{Calculation of path length, Bjorken energy density and fractional momentum loss}

Table~\ref{tab:table1} shows the different data and the inelastic nucleon-nucleon cross sections which were used to extract the quantities listed below. 

\begin{table}[b]
\caption{\label{tab:table1}%
The inelastic nucleon-nucleon cross section for the different systems considered in this work. 
}
\begin{ruledtabular}
\begin{tabular}{llc}
\textrm{System}&
\textrm{$\sqrt{s_{\rm NN}}$ (GeV)}&
\textrm{$\sigma_{\rm NN}^{\rm inel}$ (mb)}\\
\colrule
Au-Au & 62.4~\cite{Adare:2015cua} & 36.0  \\
Au-Au & 200~\cite{Adare:2015cua} & 42.3  \\
Cu-Cu & 200~\cite{Adare:2015cua} & 42.3  \\
Pb-Pb & 2760~\cite{Adam:2015kca} & 64.0  \\
Pb-Pb & 5020~\cite{Khachatryan:2016odn} & 70.0  \\
\end{tabular}
\end{ruledtabular}
\end{table}

\begin{figure*}[t!]
\begin{center}
   \includegraphics[width=1.0\textwidth]{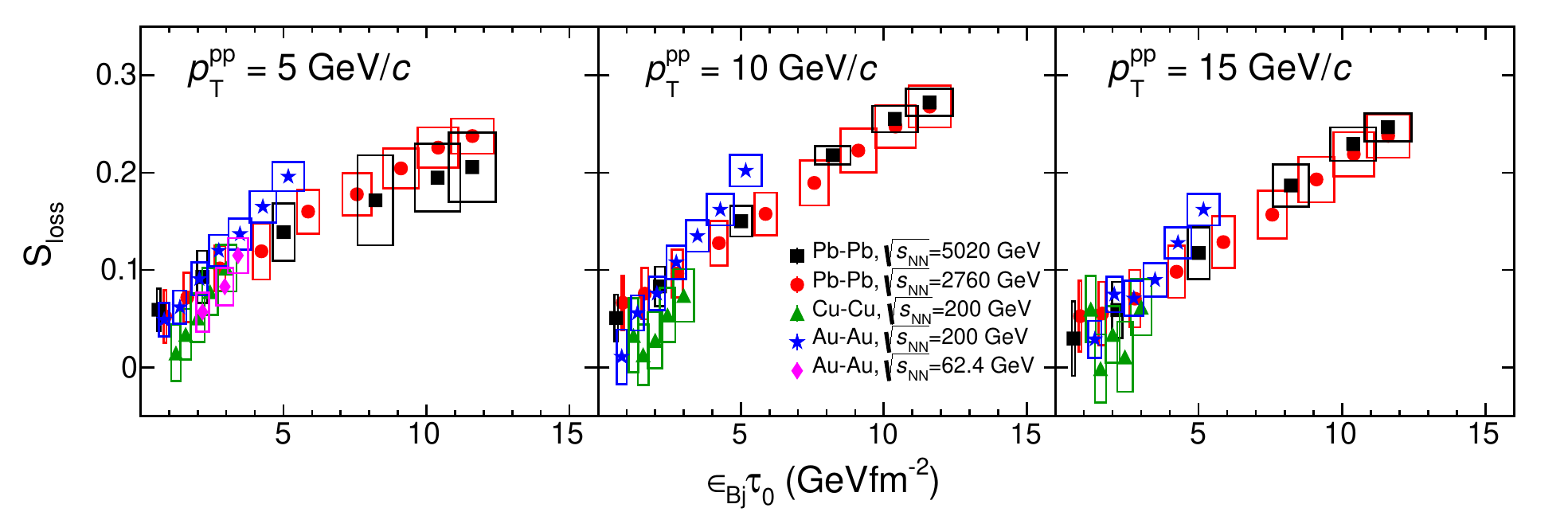}
\caption{\label{fig:2a} (Color online) Fractional momentum loss ($S_{\rm loss}$) as a function of $\epsilon_{\rm Bj}\tau_{0}$. Results for three values of transverse momentum measured in pp collisions are displayed: $p_{\rm}^{\rm pp}=5$\,GeV/$c$ (left), 10\,GeV/$c$ (middle) and 15\,GeV/$c$ (right).  Data from Pb-Pb at $\sqrt{s_{\rm NN}}=2.76$~\cite{Abelev:2012hxa} and 5.02\,TeV~\cite{Khachatryan:2016odn}, Au-Au at 62.4 and 200\,GeV,  and Cu-Cu at $\sqrt{s_{\rm NN}}=200$\,GeV~\cite{Adare:2015cua} are used for the extraction of the quantities. Systematic uncertainties are displayed as boxes around the data points.} 
\end{center}
\end{figure*}

\paragraph{Characteristic path-length.} For each colliding system (see table~\ref{tab:table1}), the nuclear overlap area was estimated from the number of participants ($N_{\rm part}$) distribution obtained from Glauber simulations~\cite{Loizides:2014vua}.  The scattering centers were randomly generated following such a distribution which is denser in the middle and decreasing toward the edge. Some examples are shown in Fig.~\ref{fig:1} which displays the distributions of the location of the scattering centers assumed for central (0-5\%), semi-central (20-30\%) and peripheral (70-80\%) Pb-Pb collisions at \snnt{2.76}. Then, for each production center the direction was determined by randomly sampling the azimuthal angle using a uniform distribution between 0 and 2$\pi$\,rad. With this information the distance from the scattering center to the edge of the area was calculated. The RMS of the distance distribution was considered as the characteristic path-length of the system ($L$). For instance, in the case of \snnt{2.76} Pb-Pb collisions the characteristic path-length ranged from 1.73\,fm to 3.13\,fm going from the most peripheral to the most central collisions, respectively. It is important to recall that the inclusion of more realistic models of initial conditions is not expected to modify the average geometrical properties~\cite{Gale:2013da}.

\paragraph{Energy density.} The Bjorken energy density~\cite{Bjorken:1982qr} is defined as
\begin{equation}
\epsilon_{\rm Bj}=\frac{1}{\tau_{0}A_{\rm T}} {\langle \frac{{\rm d} E_{\rm T}}{{\rm d}y} \rangle},
\end{equation}
where $\tau_{0}$ is the proper time when the QGP is equilibrated, $A_{\rm T}$ is the transverse area of the system and ${\langle \frac{{\rm d} E_{\rm T}}{{\rm d}y} \rangle}$ is the mean transverse energy per unit rapidity. As done it by the PHENIX Collaboration the transverse area was approximated using $\sigma_{\rm x}$ and $\sigma_{\rm y}$ being the RMS of the distributions of the $x$ and $y$ positions of the participant nucleons in the transverse plane, respectively. Moreover, since $\tau_{0}$ is model dependent, $\epsilon_{\rm Bj}\tau_{0}$ is used instead $\epsilon_{\rm Bj}$~\cite{Cuautle:2016huw}. For heavy-ion collisions at $\sqrt{s_{\rm NN}}=$ 62.4\,GeV and 200\,GeV, the $\epsilon_{\rm Bj}\tau_{0}$ values reported in~\cite{Adare:2015cua} were used. Energy density values for \snnt{2.76} Pb-Pb are also available, however, we used our own estimates and they were found to be consistent to those published in~\cite{Adare:2015cua}.

Since no transverse energy data are available for the top LHC energy ($\sqrt{s_{\rm NN}}=5.02$\,TeV), the corresponding values were extrapolated using the fact that, within 25\%, $\langle {\rm d}E_{\rm T}/{\rm d}\eta \rangle / \langle N_{\rm ch}/{\rm d}\eta \rangle$ vs. $\langle N_{\rm part} \rangle$ is nearly energy independent. This has been reported by the ALICE Collaboration, where such a scaling holds for measurements at RHIC and run I LHC energies~\cite{Adam:2016thv}. Due to this assumption, 15\% was assigned as systematic uncertainty to ${\rm d}E_{\rm T}/{\rm d}\eta$ for Pb-Pb collisions at \snnt{5.02}. In order to convert from pseudorapidity to rapidity, a factor that compensates the corresponding phase space difference is calculated. For \snnt{5.02} it amounts to 1.09 with a systematic uncertainty of 3\% like in \snnt{2.76}~\cite{Adare:2015cua}.  

\paragraph{Fractional momentum loss.}
The fractional momentum loss  ($S_{\rm loss}$) of large transverse momentum hadrons has been explored by the PHENIX Collaboration~\cite{Adare:2015cua}. Such a quantity is defined as
\begin{equation}
S_{\rm loss } \equiv  \frac{\delta \pt}{\pt^{\rm pp}} = \frac{\pt^{\rm pp}-\pt^{\rm A-A}}{\pt^{\rm pp}}
\end{equation}
where $\pt^{\rm A-A}$ is the \pt of the A-A measurement and $\pt^{\rm pp}$ is that of the pp measurement scaled by the average number of binary collisions $\langle N_{\rm coll} \rangle$ of the corresponding A-A centrality class at the same yield of the A-A measurement. The quantity is calculated as a function of $\pt^{\rm pp}$ and can be related to the original partonic momentum. Therefore, $S_{\rm loss}$  can be used to measure the parton energy loss, which should reflect the average fractional energy loss of the initial partons.

The calculation of $S_{\rm loss}$ is done as follows. The inclusive charged particle $p_{\rm T}$ spectrum in pp collisions is scaled by the $\langle N_{\rm coll} \rangle$ value corresponding to the centrality selection of the A-A measurement. Then, a power-law function is fitted to the scaled pp spectrum. And finally, the $p_{\rm T}^{\rm pp}$ corresponding to the scaled pp yield which equals the A-A yield, at the point of interest ($p_{\rm T}^{\rm A-A}$), is found using the fit to interpolate between scaled pp points.

\begin{figure*}[t!]
\begin{center}
   \includegraphics[width=1.0\textwidth]{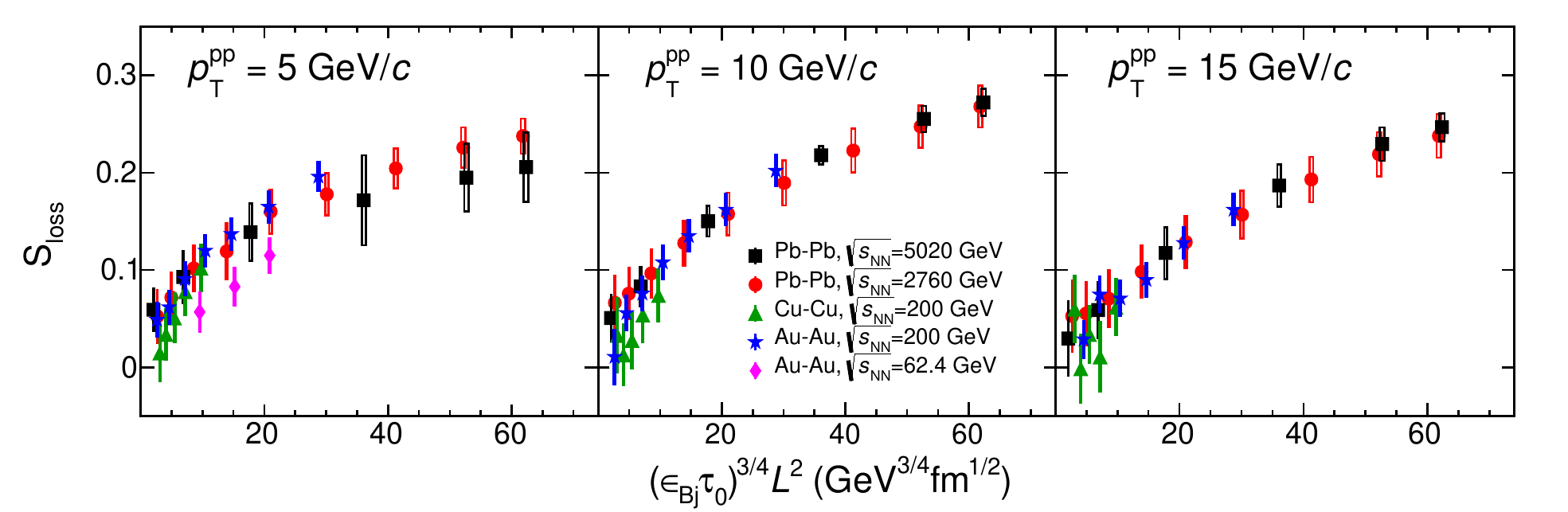} 
\caption{\label{fig:2b} (Color online) Fractional momentum loss ($S_{\rm loss}$) as a function of  $(\tau_{0}\epsilon_{\rm Bj})^{3/4}L^{2}$. Results for three values of transverse momentum measured in pp collisions are displayed: $p_{\rm}^{\rm pp}=5$\,GeV/$c$ (left), 10\,GeV/$c$ (middle) and 15\,GeV/$c$ (right).  Data from Pb-Pb at $\sqrt{s_{\rm NN}}=2.76$~\cite{Abelev:2012hxa} and 5.02\,TeV~\cite{Khachatryan:2016odn}, Au-Au at 62.4 and 200\,GeV,  and Cu-Cu at $\sqrt{s_{\rm NN}}=200$\,GeV~\cite{Adare:2015cua} are used for the extraction of the quantities. Systematic uncertainties are displayed as boxes around the data points.} 
\end{center}
\end{figure*}

The systematic uncertainties were estimated as follows. The pp (A-A) yield was moved up (down) to the corresponding edges of the systematic uncertainties, this gives the maximum deviation between both transverse momentum spectra which can be used to quantify the maximum effect on the extraction of $S_{\rm loss}$. For the most central Pb-Pb collisions at \snnt{5.02} the systematic uncertainties were $\sim17$\%, $\sim~5$\% and $\sim6$\% for $\pt^{\rm pp}=$5, 10 and 15\,GeV/$c$, respectively.
 
\section{Results and discussion}

At the LHC, it has been observed that the effects attributed to flow and new hadronization mechanisms like recombination, if any, are only relevant for transverse momentum below 10\,GeV/$c$~\cite{Abelev:2014laa,Adam:2015kca}. Therefore, previous data-driven studies of path-length dependence of  parton energy loss obtained using the elliptic flow coefficient ($v_{2}$) measurements could only provide results for $\pt>10$\,GeV/$c$~\cite{Christiansen:2013hya}. Because for high \pt, $v_{2}$ is expected to be entirely attributed to jet quenching reflecting the azimuthal asymmetry of the path-length~\cite{Christiansen:2016uaq}. However, for jet quenching phenomenology it is also important to explore the intermediate \pt (5-10\,GeV/$c$), even if the aforementioned effects (e.g. flow) are present. Since the present work does not rely on $v_{2}$ measurements, $S_{\rm loss}$ can be studied starting from  $\pt^{\rm pp}=5$\,GeV/$c$.

Figure~\ref{fig:2a} shows the fractional momentum loss as a function of $\epsilon_{\rm Bj}\tau_{0}$ for three different $p_{\rm T}^{\rm pp}$ values: 5\,GeV/$c$ (left), 10\,GeV/$c$ (middle) and 15\,GeV/$c$ (right). For \pt larger than 10\,GeV/$c$ the fractional momentum loss increases linearly with  energy density. However, the rise of $S_{\rm loss}$ with $\epsilon_{\rm Bj}\tau_{0}$ seems to be steeper at RHIC than at LHC energies. For transverse momentum of 5\,GeV/$c$, $S_{\rm loss} \propto \epsilon_{\rm Bj}\tau_{0}$ is not valid anymore. Therefore, the universality of $S_{\rm loss}$ vs. $\epsilon_{\rm Bj}\tau_{0}$ reported in~\cite{Adare:2015cua} is hard to argue. It is worth noticing that the PHENIX Collaboration reported $S_{\rm loss}$ in logarithmic scale, therefore the differences (which are pointed out here) were not obvious.

Now, the present study explores potential scaling properties of $S_{\rm loss}$ with energy density and path-length. For this, Fig.~\ref{fig:2b} shows the dependence of $S_{\rm loss}$ with $(\epsilon_{\rm Bj}\tau_{0})^{3/4}L^{2}$. The phenomenological motivation of using this variable has been already discussed in the introduction. In contrast with the previous case, the increase of $S_{\rm loss}$ is not linear. Moreover, as highlighted in~\cite{Christiansen:2013hya} a weak point of this representation is that the extrapolation to $(\epsilon_{\rm Bj}\tau_{0})^{3/4}L^{2}=0$ does not give a parton energy loss equal to zero. Though, the data from the different energies follow the same trend, which in principle can be attributed to the quadratic path-length which was introduced.

\begin{figure*}[t!]
\begin{center}
    \includegraphics[width=1.0\textwidth]{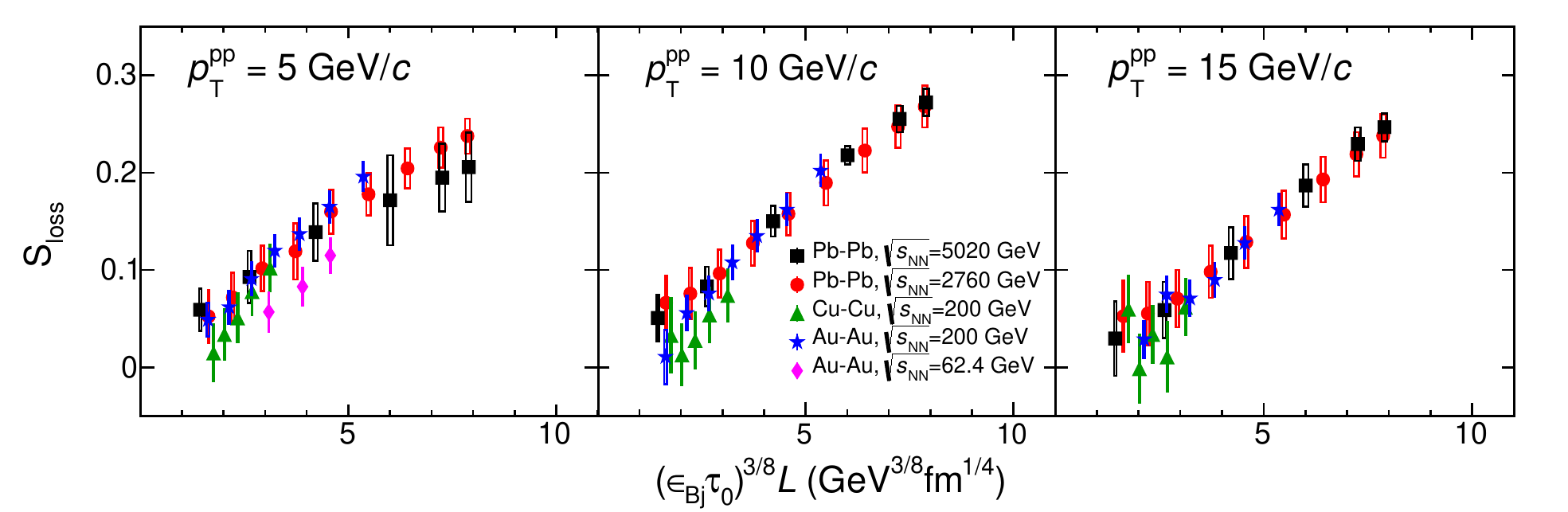}    
    \includegraphics[width=1.0\textwidth]{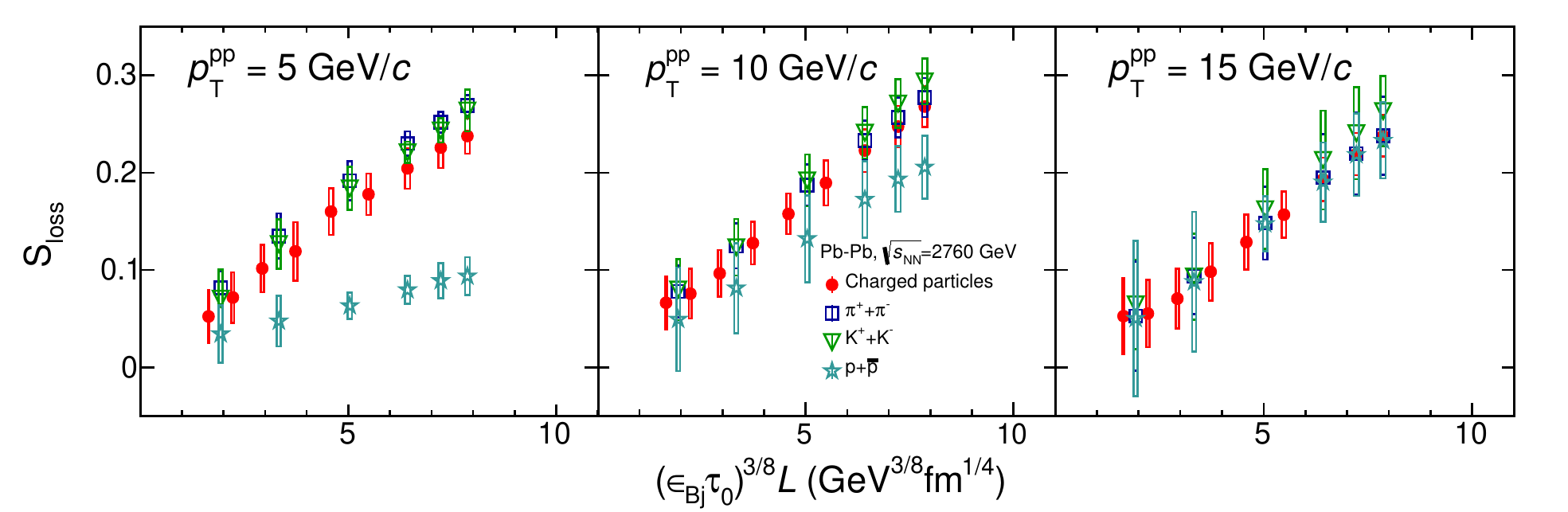}    
\caption{\label{fig:3} (Color online) Fractional momentum loss ($S_{\rm loss}$) as a function of  $(\tau_{0}\epsilon_{\rm Bj})^{3/8}L$. Results for three values of transverse momentum measured in pp collisions are displayed: $p_{\rm}^{\rm pp}=5$\,GeV/$c$ (left), 10\,GeV/$c$ (middle) and 15\,GeV/$c$ (right).  Data from Pb-Pb at $\sqrt{s_{\rm NN}}=2.76$~\cite{Abelev:2012hxa,Adam:2015kca} and 5.02\,TeV~\cite{Khachatryan:2016odn}, Au-Au at 62.4 and 200\,GeV,  and Cu-Cu at $\sqrt{s_{\rm NN}}=200$\,GeV~\cite{Adare:2015cua} are used for the extraction of the quantities. Systematic uncertainties are displayed as boxes around the data points. Results for inclusive charged particles measured at different $\sqrt{s_{\rm NN}}$ are displayed in the top panel. The bottom panel shows the results for charged pions, kaons and (anti)protons in Pb-Pb collisions at \snnt{2.76}.} 
\end{center}
\end{figure*}

The top panel of Fig.~\ref{fig:3} shows the $(\epsilon_{\rm Bj}\tau_{0})^{3/8}L$ dependence of the fractional momentum loss. Within uncertainties, $S_{\rm loss}$ increases linearly with $(\epsilon_{\rm Bj}\tau_{0})^{3/8}L$ for all the $\pt^{\rm pp}$ values which were explored. Moreover, the functional form of $S_{\rm loss}((\epsilon_{\rm Bj}\tau_{0})^{3/8}L)$ seems to be the same for all the systems which are considered.  This is the first time in which an universal scaling of $S_{\rm loss}$ vs. $(\epsilon_{\rm Bj}\tau_{0})^{3/8}L$ is observed for a broad interval of energies ranking from 62.4 up to 5020\,GeV. It is important to mention that recent studies combining $R_{\rm AA}$ and $v_{n}$ at
high $p_{\rm T}$ in realistic hydrodynamics plus jet quenching simulations seem to favor a linear path length dependence of energy loss~\cite{Noronha-Hostler:2016eow,Betz:2016ayq}. Another important observation is that $S_{\rm loss}$ exhibits an overall decrease going from $p_{\rm T}^{\rm pp}=10$ to $p_{\rm T}^{\rm pp}=15$\,GeV/$c$ which amounts to $\sim20$\%. This is consistent with the expected behavior at high $p_{\rm T}$: $S_{\rm loss} (\sim \delta_{p_{\rm T}}/p_{\rm T}) \propto 1/\sqrt{p_{\rm T}}$ which is in agreement with the observation that $R_{\rm AA}$ tends to unity at very high $p_{\rm T}$~\cite{Khachatryan:2016odn}.

It is worth noting that for $\pt^{\rm pp}=5$\,GeV/$c$ a subtle change in the slope is observed at $(\epsilon_{\rm Bj}\tau_{0})^{3/8}L\sim 4$\,GeV$^{3/8}$fm$^{1/4}$, because there other medium effects like flow could be relevant. Actually, only for the corresponding centrality class (0-40\%) the average transverse momentum for different particle species was found to scale with the hadron mass~\cite{Ortiz:2015cma}. Moreover, it is well know that at intermediate \pt (2-10\,GeV/$c$) the baryon-to-meson ratio in heavy-ion collisions is higher than that in pp collisions~\cite{Abelev:2014laa,Adam:2015kca}. In order to study the particle species dependence of $S_{\rm loss}$, the bottom panel of Fig.~\ref{fig:3} shows the charged pion, kaon and (anti)proton $S_{\rm loss}$ as a function of $(\epsilon_{\rm Bj}\tau_{0})^{3/8}L$ for Pb-Pb collisions at \snnt{2.76}. Within uncertainties, for $\pt^{\rm pp}\geq10$\,GeV/$c$ the functional form of $S_{\rm loss}$ is the same for the different identified particles and consistent with that measured for inclusive charged particles. While, for $\pt^{\rm pp}=5$\,GeV/$c$, the functional form of $S_{\rm loss}$ is only the same for inclusive charged particles, pions and kaons. Albeit the slope of the increase is significantly reduced for (anti)protons, $S_{\rm loss}$ is still observed to increase linearly  with $(\epsilon_{\rm Bj}\tau_{0})^{3/8}L$. Therefore, the change in the particle composition at $\pt<10$\,GeV/$c$ for 0-40\% Pb-Pb collisions could cause the subtle change in the slope at $(\epsilon_{\rm Bj}\tau_{0})^{3/8}L\sim 4$\,GeV$^{3/8}$fm$^{1/4}$ observed in the top panel of Fig.~\ref{fig:3}.


Last but not least, it is important to point out that assuming $\epsilon_{\rm Bj}\tau_{0}$ between 0.2 and 1.2\,GeV/fm$^{2}$ as calculated in the string percolation model for p-Pb collisions at 5.02\,TeV~\cite{2016chep.confE1152B}. Or $\epsilon_{\rm Bj}\tau_{0}\sim0.641$\,GeV/fm$^{2}$ which has been extracted from minimum bias pp collisions at $\sqrt{s}=$7 and 8\,TeV~\cite{Csanad:2016add}. One would expect jet quenching in p-Pb collisions, albeit, the size of the effect would be rather small  for $\pt^{\rm pp}>10$\,GeV/$c$ ($0<S_{\rm loss}<0.05$) compared with the large one predicted for pp collisions by some models~\cite{Zakharov:2013gya}. However, within the current systematic uncertainties reported for p-Pb collisions at \snnt{5.02} it is hard to draw a conclusion based on data~\cite{Adam:2016dau,Khachatryan:2016odn,Aad:2016zif}. The results suggest the importance of studying how different QGP-related observables evolve as a function of quantities like energy density which is crucial to understand the similarities between pp and AA collisions~\cite{Paic:2016jep}.

\section{Conclusions}

The inclusive charged particle production in heavy-ion collisions at $\sqrt{s_{\rm NN}}=62.4$ and 200\,GeV (2.76 and 5.02 TeV) measured by experiments at the RHIC (LHC) were used to extract the fractional momentum loss ($S_{\rm loss}$) and the Bjorken energy density. Using MC Glauber simulations, a characteristic path-length was estimated for the different collision centralities and for each colliding system. Surprisingly,  for all the transverse momentum values which were explored: $5<\pt<20$\,GeV/$c$, $S_{\rm loss}$ was found to increase linearly with $(\epsilon_{\rm Bj}\tau_{0})^{3/8}L$ being $\tau_{0}$ the equilibration time. Moreover, an universal functional form was found to describe the data from the different colliding systems which were analysed. In contrast, this universal (linear) behaviour is not observed if the scaling variable $(\epsilon_{\rm Bj}\tau_{0})^{3/8}L$ is replaced by $\epsilon_{\rm Bj}\tau_{0}$ ($(\epsilon_{\rm Bj}\tau_{0})^{3/4}L^{2}$). The linear increase of $S_{\rm loss}$ is also observed for identified charged particles (pions, kaons and protons) even for $\pt^{\rm pp}=5$\,GeV/$c$. The behaviour of data could provide additional constraints to phenomenological models of jet quenching not only for heavy-ion collisions but also in the jet quenching searches in small collisions systems.

\begin{acknowledgments}
Support for this work has been received from CONACyT under the grant number 280362 and PAPIIT-UNAM under Project No. IN102118.
\end{acknowledgments}



\bibliography{mybibfile}
\end{document}